\documentclass[12pt]{article}
\usepackage{fullpage}
\usepackage{hyperref}
\def\lesim{\ \hbox to 0 pt{\raise .6ex\hbox{$<$}} \lower
  .5ex\hbox{$\sim$}\ }
\def\dld{d\lambda{}}
\def\ld{\lambda{}}

\title{Statistical Uncertainty in Line Shift and Width Interpretation}
\author{I.H.Hutchinson}
\date{Plasma Science and Fusion Center\\ MIT, Cambridge, MA, USA}

\begin{document}
\maketitle
\begin{abstract}
Elementary but general statistical analyses determine the uncertainty
arising from photon statistics in measuring a line shift and
width. Account is taken of a background as well as the required
signal.
\end{abstract}

\section{Introduction}

In many spectroscopic situations, the shift and width of a spectral
line is desired. The shift is most often wanted as a measure of the
mean velocity of the emitters in the line of sight, and the width as
their Doppler temperature. The statistical uncertainty of such a
measurement can be written down extremely simply, provided the problem
is approached in the right way.

If there is present, in addition to the line signal of interest, a
background signal, which is to be subtracted from the signal, this
complicates the problem somewhat. Nevertheless, a simple result is
obtained. The way that the instrumental spectral line shape affects
the result can also be trivially included. 

Although the statistical analysis set forth here is elementary, it
does not appear in a usable form, to the author's knowledge, in the
standard textbook discussions of uncertainty. Nor are most students or
researchers familiar with it. Therefore it seems useful to set
it forth in a pedagogic style. The required mathematical results will
be cited from an introductory text book\cite{meyer66} as references to
its numbered theorems. 

Classical statistics (uncorrelated photons) rather than Bose-Einstein
statistics will be assumed throughout, as is an excellent
approximation in most experimental situations.

\section{Line Shift}

\subsection{Perfect Spectrometer, Zero Background}

We suppose initially that a perfect spectrometer is available, which
measures the exact wavelength $\lambda$ of every photon arriving. The
line shape is taken to have the form $I(\lambda)$, where $I$ is the
probability distribution function of photon wavelength. That is, the
probability of any photon having wavelength in the range
$(\lambda,\lambda+d\lambda)$, in the limit of small $\dld$, is
$I(\ld)\dld$. 

A measurement consists of collecting photons for a certain time
interval. We shall approximate this as being to collect a fixed
\emph{total number} of photons. Provided the number of photons in the
sample, $N$, is large, this will introduce negligible error, at least
in the line position. The formulation of our problem of determining
the uncertainty in a line-shift measurement is then to determine what
is the probability distribution (or at least its width) for the line
position deduced from the sample of $N$ independent photons. There
are, of course, many complicated possible ways of determining the line
position of the photon sample. However, provided that these methods
are not able to take advantage of important \emph{a priori}
information about line shifts, they will have no intrinsic advantage
over the simplest measure of line position, namely the \emph{centroid}
of the photon wavelength distribution:
\begin{equation}\label{nobackgnd}
\mu_k = {1\over N}\sum_{i=1}^N \ld_i .
\end{equation}
The index $k$ here is used to refer to the $k$th measurement (or
sample), and $i$ refers to the photon number within that sample.

The advantage of using the centroid as the line position metric, is
that the distribution of $\mu_k$ is the subject of the Central Limit
Theorem of statistics (\cite{meyer66} theorem 12.1), which states that
if the standard deviation of the population $I(\ld)$ from which the
sample is drawn is $\sigma_I$, then in the limit of large $N$, the
values of $\mu$ are distributed with normal (Gaussian) distribution,
with standard deviation
\begin{equation}\label{mudist}
\sigma_\mu = \sigma_i/\sqrt{N} .
\end{equation}
Note that this result is regardless of the shape of the original
distribution  $I(\ld)$. 

Equation (\ref{mudist}) gives the required uncertainty in the measured
line-shift. It is the width of the line to be measured divided by the
square root of the number of photons in the sample. Moreover the
Central Limit Theorem also tells us the bonus that the distribution of
$\mu$ is Gaussian.

\subsection{Perfect Spectrometer, Background Subtraction}

Now we consider a situation in which we know that there are, in
addition to the photons we care about, background photons, which we
will have to subtract from our spectrum. Suppose the background
spectrum to be given by a probability distribution $B(\ld)$ (so that
$\int B(\ld)\dld =1$), having a standard deviation (spectral width)
$\sigma_B$ and a centroid $\mu_B$.

A measurement consists of obtaining a large number of photons, each of
whose wavelength is measured. Of those photons, $N_I$ are the signal
photons, and $N_B$ are the background photons, $N=N_I+N_B$. We don't
know which photon is which, but for purposes of reference, we will
suppose them to be ordered such that the first $N_I$ are the signal
photons. We will assume initially that the numbers $N_I$ and $N_B$
are known and fixed.

Our best estimate of the background is that its spectral photon
density is $N_B B(\ld)$, so that the background contribution to the
centroid sum $\sum \lambda$ is on average $N_B \mu_B$. We will subtract this
background from the spectrum and then obtain the centroid of the
remaining ``signal''. Therefore the estimate we should use for the
signal photons' centroid is
\begin{equation}\label{backgnd}
\mu_{Ik} = {1\over N_I}\left(\sum_{i=1}^N \lambda_i - N_B \mu_B \right)
={1\over N_I}\left(\sum_{i=1}^{N_I} \lambda_i+\sum_{N_I+1}^N
\lambda_i - N_B \mu_B \right) .
\end{equation}
The second equality in eq (\ref{backgnd}) is written to show that if
we had a perfect knowledge of the background contribution, the last two
terms exactly cancel, returning us to our original form (\ref{nobackgnd}).

However, the two terms do not exactly cancel, even if our knowledge of
$B(\ld)$ is perfect, because of the statistics of the background
photons. We know that the centroid of the background photons in the
sample, 
\begin{equation}
\mu_{Bk} = {1\over N_B} \sum_{i=N_I+1}^N \lambda_i ,
\end{equation}
is distributed with Gaussian probability distribution with mean
$\mu_B$ and standard deviation $\sigma_B/\sqrt{N_B}$. Therefore our
line position estimate is the sum of three terms. The first is the
centroid of the signal photons in the sample, distributed with
standard deviation $\sigma_I/\sqrt{N_i}$. The second is the centroid
of the background photons in the sample, $\mu_{Bk}$ times $N_B/N_I$;
and the last is a constant, which we chose to annihilate the mean of
$\mu_{Bk}$.

Utilizing the theorem that a random variable that is the sum of two
other independent random variables has a distribution whose mean is
the sum of the means (\cite{meyer66} Property 7.4), and whose variance
(standard deviation squared) is the sum of the variances (Property
7.9), we can immediately deduce that the distribution of our line
shift estimate $\mu_k$ has mean equal to the centroid of $I(\ld)$ and
standard deviation:
\begin{equation}\label{lineuncertainty}
\sigma_\mu = \sqrt{{\sigma_I^2\over N_I} + {\sigma_B^2 N_B\over
N_I^2}}  .
\end{equation}

This important result shows how rapidly a background can
come to dominate the uncertainty in the measurement of line shift. If
$\sigma_I \sim \sigma_B$, then as soon as $N_B > N_I$ it is the second
term, coming from statistics of the background photons, that
determines the shift uncertainty. 

Notice also, that if the background spectrum is broad, for example
flat, in the vicinity of the signal line, then the effective
$\sigma_B$ is determined purely by the spectral band chosen over which
to sum the photons. To minimize this width we should choose a spectral
band that is just wide enough to encompass the signal line, but no
wider. In that case, the background width will be wider than the
actual line spread by a roughly factor of 2 or 3, depending on the
exact band width chosen.

In the contrasting case where the background line mimics the signal
line in width so that $\sigma_I=\sigma_B$, 
the resultant uncertainty is 
\begin{equation}
\sigma_\mu  = {\sigma_B \over \sqrt{N_I}} \sqrt{1+N_B/N_I} .
\end{equation}
Either way, when $N_B\gg N_I$, the uncertainty becomes 
\begin{equation}
\sigma_\mu \approx {\sigma_B \sqrt{N_B} \over N_I} .
\end{equation}

Since both of the random contributions to $\mu_k$ are normally
distributed (by the Central Limit Theorem), and the distribution of a
sum of normally distributed random variables is normally distributed
(\cite{meyer66} Theorem 10.6),
$\mu_k$ itself is also normally distributed.

\subsection{Offset Background: Background-fraction Fluctuations.}

If the background distribution is substantially shifted in its mean
from the mean of the signal, an additional effect may be non-negligible.
We must account for 
the statistical uncertainty in the fraction of sample photons
that is actually background, rather than signal. In other
words, although we will assume we know the average number of signal
and background photons, $N_I$ and $N_B$, we do not know the actual
numbers for any specific sample, $N_{Ik}$ and $N_{Bk}$. Therefore, our
expression for the estimate of the signal mean, eq (\ref{backgnd}),
must more precisely be written

\begin{equation}\label{fracbackgnd}
\mu_{Ik} = {1\over N_I}\left(\sum_{i=1}^N \lambda_i - N_B \mu_B \right)
={1\over N_I}\left(\sum_{i=1}^{N_{Ik}} \lambda_{Ii}+\sum_{N_{Ik}+1}^N
\lambda_{Bi} - N_B \mu_B \right) ,
\end{equation}
where we explicitly denote with a subscript those photons that are
signal and background. Grouping terms appropriately we then write
\begin{equation}
\mu_{Ik}
={1\over N_I}
\left(\sum_{i=1}^{N_{I}}\lambda_{Ii} + \sum_{i=N_I+1}^{N_{Ik}}\lambda_{Ii}
-\sum_{N_{I}+1}^{N_{Ik}}\lambda_{Bi}+ \sum_{N_{I}+1}^N
\lambda_{Bi} - N_B \mu_B \right) ,
\end{equation}
which is precisely the previous expression, eq (\ref{backgnd}),
\emph{plus} the combination
\begin{equation}
{1\over N_I}\left(
  \sum_{i=N_I+1}^{N_{Ik}}\lambda_{Ii}
-\sum_{N_{I}+1}^{N_{Ik}}\lambda_{Bi}\right) .
\end{equation}
This expression is zero if $N_{Ik}=N_{I}$, but for fixed difference
$N_{Ik}-N_I$ has expectation $(N_{Ik}-N_I)(\mu_I-\mu_b)/N_I$.  To
lowest order in $1/N$, this additional term is therefore distributed
as $\mu_I-\mu_B$ times the fractional deviation of $N_{Ik}$ from
$N_I$. The
standard deviation of $N_{Ik}/N$ is
$\sqrt{N_I N_B /N^3}$. So the additional term has standard
deviation $(\mu_I-\mu_b)\sqrt{ N_B/(N_I N)}$.


Consequently the standard deviation of the estimate of the line
position (eq \ref{lineuncertainty}) is generalized to
\begin{equation}\label{linegen}
\sigma_\mu = \sqrt{{\sigma_I^2\over N_I} +
{(\mu_I-\mu_B)^2N_B\over N_I N} +
 {\sigma_B^2 N_B\over N_I^2}}  .
\end{equation}
Recalling the considerations in the previous subsection,
the wavelength collection region must be chosen approximately centered
on $\mu_I$ and with (half) width no more than 2-3 times $\sigma_I$. As a
result the mean collected background photon wavelength cannot deviate
more than $2-3 \sigma_I$ from $\mu_I$: $|\mu_I-\mu_B|\lesim
2\sigma_I$. The additional (center) term in eq (\ref{linegen}) 
is  most significant when
$N_B\sim N_I$ because if $N_B\ll N_I$, it is negligible compared with
the first term and if $N_I\ll N_B$ it is negligible compared with the last.
Thus the worst case is that the
extra term increases the uncertainty arising from the signal width
term by about a factor of 2. More typically,  $|\mu_I-\mu_B| <
\sigma_I$, in which
case the additional term is essentially negligible.

\subsection{Imperfect Spectrometer or Complex Line Shape}

In practice a spectrometer has a finite instrumental line width. Its
effects can often be described by an instrument line shape $S(\ld)$, such
that the observed spectrum is the \emph{convolution} of $S(\ld)$ with
the incident spectrum. Consequently a perfectly monochromatic line, $I(\ld)=\delta(\ld-\ld_0)$,
acquires the instrumental shape $S(\ld-\ld_0)$. Because the instrument shape can be
taken to have zero centroid shift, the convolution introduces no
systematic shift of the line, nor, more importantly, any random
shift. What it does, however, is to broaden the width of the observed
line, relative to what would otherwise have been observed. 

From the point of view of statistics, all of the previous arguments
still apply. But they apply to the \emph{instrumentally broadened}
observed line rather than to the original line shapes. In particular
all of the foregoing formulas will still apply if the $\sigma$ factors
are taken as the widths of the lines after convolving with the
instrumental function. In other words, if the observed widths are
used. 

By the same token, if the lines being examined have complex line shape
due, for example, to a multiplet structure, then all of the foregoing
formulas \emph{still} apply. But the width of the lines is then
perhaps dominated by the multiplet structure rather than the other
broadening mechanisms.

\subsection{Finite-width Wavelength Bins}

A situation of importance for modern spectrometers, is when the
instrument does \emph{not} simply produce a convolution of the line
with an instrumental line-shape. Instead, one might have a series of
bins of finite-width into which the photons are sorted. This will
occur, for example, with a multi-element detector when each of the
different elements corresponds to a different wavelength bin. It is
clear that if the bin width is much less than the observed width of
the line, then the binning will amount simply to a discrete
approximation to the integrals and sums invoked above.

However, if the bin size is larger than the line width, then strongly
non-linear effects on the shift will be introduced. For example, as a
narrow line moves across a wider bin, no detectable change in the
spectrum occurs until the line approaches the boundary between
bins. Then as it crosses the bin boundary, a step in apparent position
of the photons occurs from one bin to the next. 

If the bins are of equal width, perhaps the best way to see the effect
of the bins is as follows. We deduce the centroid by assigning each
photon in a bin a wavelength equal to the center of the bin; this is the
most reasonable unbiassed estimate. Let the center of the $j^{th}$ bin be at
wavelength $\ld_j = j \Delta\ld$ (measured from a convenient zero). Then
the error in the centroid for a photon at position $\ld$ lying in the
$j^{th}$ bin is $\ld-\ld_j$, which is a linear ramp. If the photon
lies in the next bin $j+1$, the error has an identical ramp
($\ld-\ld_{j+1}$). Therefore the error function has the form of a
\emph{sawtooth}:
\begin{equation}
{\cal E}(\ld) = \ld -\ld_j ,\qquad 
\mbox{for } (\ld_j+\ld_{j-1})/2 < \ld \le (\ld_{j+1}+\ld_j)/2\ .
\end{equation}
And a line shape $I(\ld)$ acquires an error
\begin{equation}\label{binerror}
\int I(\ld) {\cal E}(\ld) \dld \ .
\end{equation}

Now ${\cal E}$ is a periodic function, and can thus be expressed as a
Fourier series
\begin{equation}
{\cal E} = \sum_{m=1}^\infty A_m \sin(2\pi m \ld /\Delta\ld)\ .
\end{equation}
When this is substituted into eq (\ref{binerror}),
we see immediately that the error arising from a line shape
$I(\ld)$ can be written as a weighted sum of the Fourier transform of
$I(\ld)$ evaluated at the (spatial) frequencies $2\pi m /\Delta\ld$. The most
important error will arise from the fundamental $m=1$, both because
$A_m$ is a decreasing function of $m$ and because the Fourier
transform will be a decreasing function of frequency. Indeed if
$I(\ld)$ is a Gaussian, then its Fourier transform is also a Gaussian,
whose width is $\sim 1/\sigma_I$. Provided that  $2\pi /\Delta\ld$
is substantially larger than this width, the binning error will become
negligibly small. This condition is equivalent to the requirement that
there be more than a few bins covering the width of the line. 

Thus, somewhat counter-intuitively, if we have a binned spectrum, it
is advantageous, from the viewpoint of line position measurement, that
the effective line-shape (the convolution of the received
line with the instrument function) should be significantly
\emph{broader} than the bin width. It is actually a substantial
\emph{disadvantage} to have a line narrower than approximately the bin
width. For that reason, a detector array ought always to be spaced
closer than the instrumental resolution.

Moreover, the naive idea that the bin size represents a minimum
resolution, and thus a minimum line-shift resolution, is
\emph{false}. For a Gaussian line, the Fourier transform decays so
rapidly at a few times its width that it becomes completely
negligible. When that occurs, if there are enough photons in the line,
the uncertainty given by eq (\ref{lineuncertainty}) may be far smaller
than the bin width.

\section{Line Width}

\subsection{Line Width, Zero Background}

Generally speaking, the line width can be analysed in a way comparable to
the shift. However, the elementary statistical theorems are
not quite as general. 

The line width can be defined in terms of the second moment of the
sample. The usual unbiassed statistical estimate for the
width ($\sigma_I$) of the population, based on a sample of size $N$,
is $S_k$, the square root of the sample variance given by:
\begin{equation}
  S_k^2 = {1\over N-1} \sum_{i=1}^N (\ld_i - \mu_k)^2
\end{equation}
If $I(\ld)$ is a Gaussian distribution, then by Theorem
10.8\cite{meyer66}, the
variable $u=S^2 (N-1)/\sigma_I^2$, which is the sum of $N$ independent
normally distributed variables, has a chi-squared distribution of
$N$ degrees of freedom. In other words, its probability
distribution function is
\begin{equation}
  {1\over 2^{N/2}}\Gamma(N/2)u^{N/2-1}e^{-u/2} 
\end{equation}
In particular, since that chi-squared distribution has mean $N$ and
variance $2N$, the variance of $S^2$ is $2N\times
[\sigma_I^2/N]^2 = \sigma_I^4 2/N$, and the standard deviation
of $S^2$ is $\sigma_I^2 \sqrt{2/N}$. 

For large $N$ (by \cite{meyer66} Theorem 9.2) if $u$ is distributed as a
chi-squared distribution of $N$ degrees of freedom, then the variable
$\sqrt{2u}$ has approximately a normal distribution of mean
$\sqrt{2N-1}$ and variance unity. Thus for large $N$ the distribution
of $S$ is Gaussian with mean
\begin{equation}
\mu_S = \sigma_I\sqrt{2N-1\over 2(N-1)}  \approx \sigma_I\ ,
\end{equation}
and standard deviation
\begin{equation}\label{widtherror}
\sigma_S ={\sigma_I\over\sqrt{2(N-1)}} \ .
\end{equation}

The value $\sigma_S$, eq (\ref{widtherror}), can be considered the
uncertainty in determination of the line width, when the width is
evaluated by using the second moment of the line shape. As before, its
should be applied to the observed line width (including instrumental
broadening). 

\subsection{Line Width, Background Subtraction}

The natural unbiassed estimate of the line width in the presence of
background photons is $S_k$, where
\begin{equation}
  S^2_k = {1\over N_I-1}\left[
\sum_{i=1}^N (\ld_i-\mu_{Ik})^2  - N_B \{\sigma_B^2+(\mu_{Ik}-\mu_B)^2\}\right] 
\end{equation}
where $\mu_{Ik}$ is given by eq (\ref{backgnd}). This can be written
\begin{eqnarray}
  S^2_k &=& {1\over N_I-1}\Big[
\sum_{i=1}^{N_I} (\ld_i-\mu_{Ik})^2+
\sum_{N_I+1}^N (\ld_i-\mu_{B})^2 - \sum_{N_I+1}^N 
2\ld_i(\mu_{Ik}-\mu_{B})
+N_B(\mu_{Ik}^2-\mu_{B}^2)
\nonumber\\
 &&\qquad - N_B \{\sigma_B^2+(\mu_{Ik}-\mu_B)^2\}\Big] 
\\
 &=& {1\over N_I-1}\Big[
\sum_{i=1}^{N_I} (\ld_i-\mu_{Ik})^2+
\sum_{N_I+1}^N\{ (\ld_i-\mu_{B})^2 -\sigma_B^2\}
+2(\mu_{B}-\mu_{Ik})\sum_{N_I+1}^N (\ld_i - \mu_B) \Big]\nonumber
\end{eqnarray}
The second and third sums in this expression come from the statistics
of the background photons. The variances of these two expressions are
given by:
\begin{equation}
  Var({1\over N_B-1}\sum_{N_I+1}^N\{ (\ld_i-\mu_{B})^2 -\sigma_B^2\})=
{2\sigma_B^4\over N_B-1}
\end{equation}
and
\begin{equation}
  Var({2(\mu_{B}-\mu_{Ik})\over N_B-1}\sum_{N_I+1}^N (\ld_i - \mu_B))=
{4(\mu_B-\mu_{Ik})^2\sigma_B^2\over N_B-1} .
\end{equation}
Assuming that the variance of $S^2$ can be written as the sum of the
variances of these three terms (which is not obvious since they aren't
independent) we get
\begin{equation}
  Var(S^2_k) = {2\sigma_I^4\over N_I-1} + 
\left(
{2\sigma_B^4\over N_B-1}
+ {4(\mu_B-\mu_{Ik})^2\sigma_B^2\over N_B-1}
\right)
\left({N_B-1\over N_I-1}\right)^2
\end{equation}

Let us drop the irrelevant small distinction between $N-1$ and $N$. Then
this expression shows that the background terms will dominate if
$(N_B/N_I)\sigma_B^4 > \sigma_I^4$. The third term will give
an additional contribution unless we can arrange that the background
spectrum is centered on the signal line to better than $\sigma_B$; but
that should be easy to accomplish; so we will ignore the third term.
Then
\begin{equation}\label{vars2}
  Var(S^2_k) = {2\sigma_I^4\over N_I} + {2\sigma_B^4N_B\over N_I^2}
\end{equation}
And of course the expectation (mean) of $S^2$ is $\sigma_I^2$. 

Provided the distribution of $S_k^2$ is narrow, we can regard its
distribution as centered at $\mu_{S^2}$ with a much smaller
width $\sigma_{S^2}$. Then the square-root variable $S$, can be
expanded schematically as 
\begin{equation}
  S = \sqrt{S^2} = \sqrt{\mu_{S^2}+\sigma_{S^2}}\approx 
 \sqrt{\mu_{S^2}}(1+ {\textstyle {1\over2}}\sigma_{S^2}/\mu_{S^2})
\end{equation}
which shows that the mean of $S$ is $\sqrt{\mu_{S^2}}$ but the
standard deviation of $S$ is $\approx{1\over2}\sigma_{S^2}/\sqrt{\mu_{S^2}}$.
Applying this to the result of eq (\ref{vars2}), we find that the
standard deviation of our estimate $S$ of the line width is
\begin{equation}\label{widthbackgnd}
  \sigma_S \approx {1\over2\sigma_I}\sqrt{{2\sigma_I^4\over N_I} +
  {2\sigma_B^4N_B\over N_I^2}} = {\sigma_I\over\sqrt{2N_I}} \sqrt{1 +
  {\sigma_B^4 N_B\over \sigma_I^4 N_I}}.
\end{equation}
This result is consistent with eq (\ref{widtherror}), which was
obtained using more specific distribution assumptions, but without the
approximations made here. 

\section{Summary}

When line shift and width are measured using a large sample of $N_I$ signal
photons from a line whose width is $\sigma_I$ in the presence of $N_B$
background photons from a population centered on the signal line but
with width $\sigma_B$, the photon statistical uncertainties in the
shift and width of the signal line, deduced from the moments of the
distribution, are respectively:
\begin{equation}
\sigma_\mu  = {\sigma_I \over \sqrt{N_I}} \sqrt{1+{\sigma_B^2 N_B\over
    \sigma_I^2 N_I}}   
\end{equation}
and
\begin{equation}
\sigma_S= {\sigma_I\over\sqrt{2N_I}} \sqrt{1 +
  {\sigma_B^4 N_B\over \sigma_I^4 N_I}}.
\end{equation}
In these expressions, the observed widths of the line and background,
including instrumental or multiplet broadening effects, should be used.

\bibliography{probability}

\begin{thebibliography}{1}

\bibitem{meyer66}
Paul~L Meyer.
\newblock {\em Introductory Probability and Statistical Applications}.
\newblock Addison Wesley, Reading, MA, 1966.

\end{thebibliography}

\end{document}